\documentclass[runningheads,citeauthoryear]{apinv}
\usepackage{epsfig,cite,graphics}
\usepackage[T2A]{fontenc}
\usepackage[cp1251]{inputenc}

\begin{document}

\title{New light curve solution of V568~Peg and first
determination of its fundamental parameters}
\titlerunning{New light curve solution of V568~Peg}
\author{Diana Kjurkchieva\inst{1} and Sunay Ibryamov\inst{1}}
\authorrunning{D. Kjurkchieva and S. Ibryamov}
\tocauthor{D. Kjurkchieva}
\institute{Department of Physics and Astronomy, Shumen University,
115 Universitetska, 9700 Shumen, Bulgaria
 \newline
    \email{d.kyurkchieva@shu.bg};
     \email{s.ibryamov@shu.bg};
 }
\papertype{Submitted on xx.xx.xxxx; Accepted on xx.xx.xxxx}
\maketitle

\begin{abstract}
We present $V, R_c$ photometric observations of the short-period W
UMa star V568 Peg. They allowed us to improve its period. The
light curve solution revealed that V568 Peg is an overcontact
binary of A subtype with moderate fill-out factor. Its components
are K stars which undergo partial eclipses. The mass ratio was
estimated by $q$-search analysis. We established existing of big
cool spot on the primary component with almost the same parameters
during the last 4 years. Based on our light curve solution and the
\emph{GAIA} distance we calculated at the first time the masses,
radii and luminosities of the components of V568 Peg.
\end{abstract}


\section*{1. Introduction}

The temperature difference of the components of W UMa binaries
usually are around 100--300 K (only those of B subtype systems are
above 1000 K) while their masses and radii may differ considerably
(Binnendijk 1965, Lucy $\&$ Wilson 1979, Csizmadia $\&$ Klagyivik
2004). The model of Lucy (1968a, 1968b) explained this effect by a
common convective photosphere which embedded two stars near or
just above the Main Sequence. However, until now there is not a
satisfactory explanation of the mechanism of energy transfer, the
W phenomenon (the hotter component is the smaller star) and the
internal structure of the W UMa binaries. Their future fate is
also debatable issue: tight binary or merger (van Hamme 1982a,b;
Li et al. 2007). The solutions of these problems requires rich
statistics of well-determined global parameters of W UMa stars.
The \emph{GAIA} distances (Bailer-Jones et al. 2018) of a huge
number of eclipsing binaries provide invaluable possibility for
precise determination of their global parameters on the base of
ground-based observations.

This paper presents $V, R_c$ photometric observations of the
short-period W UMa-type system V568 Peg. It was observed 4 years
ago in Sloan $g', i'$ bands (Kjurkchieva et al. 2015, further on
Paper I). The main goal of the new observations was to determine
its fundamental parameters by the light curve solution and
\emph{GAIA} distance. This is the first study based on
observations by the 10-inch Schmidt-Cassegrain telescope MEADE
LX80 SC of the Shumen Astronomical Observatory.

\section*{2. Observations}

The photometric observations of the target were carried out on Aug
13 2018. We used CCD camera SBIG ST-10XME (2184 $\times$ 1472
pixels, 6.8 $\mu$m/pixel). Focal reducer TS Optics f/6.3 provides
increasing of the field of view from 20 $\times$ 14 arcmin with
resolution 0.55 arcsec/pix to 32 $\times$ 22 arcmin with
resolution 0.88 arcsec/pix (Kjurkchieva et al. 2018). The
exposures in $V$ and $R_c$ filter were 90 s and 60 s and the mean
photometric precision was 0.029 mag in both filters.

The photometric data were reduced by {\textsc{MaxIm DL 5}. An
aperture photometry was performed using four standard stars (Table
1) in the observed field whose coordinates were taken from the
catalogue 2MASS (Skrutskie et al. 2006) while their magnitudes
were from the NOMAD catalog (Zacharias et al. 2004).

The reduced data are accessible in the form of table as online
supplemental data and at http://astro.shu.bg/V568Peg/.

\begin{table}[htb]
  \begin{center}
  \caption{Coordinates and magnitudes of the target (V) and comparison (C) stars}
 \begin{tabular}{cccccc}
  \hline\hline
Label  & 2MASS ID & \multicolumn{1}{c}{RA} & \multicolumn{1}{c}{Dec} & \multicolumn{1}{c}{$V$} & \multicolumn{1}{c}{$R_c$} \\
 \hline
V      &    V568 Peg      & 23 08 13.01 & +33 03 03.77 & 12.960 & 12.550\\
C1     & 23081802+3259237 & 23 08 18.03 & +32 59 23.70 & 11.361 & 10.590\\
C2     & 23080030+3306284 & 23 08 00.30 & +33 06 28.41 & 13.770 & 13.420\\
C3     & 23074178+3307567 & 23 07 41.79 & +33 07 56.75 & 11.534 & 10.650 \\
C4     & 23075759+3302504 & 23 07 57.60 & +33 02 50.41 & 12.130 & 11.620\\
 \hline
 \end{tabular}
  \label{table1}
  \end{center}
\end{table}

\section*{3. Light curve solution}

We carried out the modeling of our data by the package
\textsc{PHOEBE svn} (Prsa $\&$ Zwitter 2005). The observational
data (Fig. 1) show that our target is an overcontact system and we
modelled them using the corresponding mode ''Overcontact binary
not in thermal contact''.

The values of target (weighting) temperature \textbf{$T_{m}$=$T_1
L_1/(L_1+L_2)+T_2 L_2/(L_1+L_2)$} determined by different ways are
slightly different: $T_{m}^{BV}$ = 5100 K is the value from the
dereddened index $(B-V)$ and relation of Sekiguchi $\&$ Fukugita
(2000); $T_{m}^{gi}$ = 5000 K is the value from the dereddened
index $(g'-i')$ and relation of Covey et al. (2007); $T_{m}^{JK}$
= 4850 K is the value from the 2MASS dereddened index $(J-K)$ and
relation of Cox (2000); $T_{m}^{G}$ = 4800 K is the value
determined by \emph{GAIA} DR2 (\emph{Gaia} Collaboration 2018).
The different values of target temperature may due at least
partially to the different phases of measurement of the color
indices ($T_{m}$ should be determined by measurements at
quadratures).

The interstellar reddening was estimated based on the following
considerations. The extinction $A_V$ in the V568 Peg direction is
0.221 mag according to the NED database (Schlafly $\&$ Finkbeiner
2011) and 0.134 mag according to the 3D model of Arenou et al.
(1992) for distance 253 pc (see Section 4). The extinction values
of NED refer to distance above 500--600 pc while the distance to
V568 Peg is considerably smaller. It was reasonable to reduce the
extinction in all colors of the NED database by the same factor
0.61(=0.134/0.221) as in $V$. This rule was used for estimation of
all dereddened indices.

Unfortunately, there were not spectra of V568 Peg for confident
temperature determination. This problem may be overcame by
low-dispersion spectral observations (for instance by future
low-dispersion spectrograph of the 2-m telescope at NAO Rozhen).
In our case $T_{m}$ = 4900 K was adopted as some average value of
the values determined by different ways.

We fixed the primary temperature $T_{1}$ = $T_{m}$ and searched
for best fit varying initial epoch $T_0$, secondary temperature
$T_{2}$, mass ratio $q$, inclination $i$ and potential $\Omega$.
Coefficients of gravity brightening 0.32 and reflection effect 0.5
appropriate for late stars were assumed. We used linear
limb-darkening law whose coefficients were interpolated (depending
on stellar temperatures and filters) according to the tables of
Van Hamme (1993). In order to reproduce the O'Connell effect of
around 0.065 mag (Fig. 1) we put a cool spot on the primary and
varied its parameters (longitude $\lambda$, latitude $\beta$,
angular size $\alpha$ and temperature factor $\kappa$).

The mass ratio determination of the partially-eclipsed binary V568
Peg required $q$-search analysis. For this aim we varied the mass
ratio in a wide interval, from 0.1 to 10.0. The $q$-search curve
(Fig. 2) exhibits two minima. The first one at $q$ = 0.4 is deeper
and narrower while the second one at $q$ = 4 is shallower and
wider. The $\chi^2$ value for $q$ = 4 is around 3 times bigger
than that for $q$ = 0.4. We carried out detailed investigation of
the solution around $q$ = 4 by varying of all parameters but
reached inconsiderable decreasing of $\chi^2$ (by several $\%$).
That is why we chose as input value $q$ = 0.4. Radial velocity
measurements could provide confirmation of our choice although the
spectral lines of the W UMa stars are broadened and blended that
leads to low-precise determination of the spectral mass ratio
(Frasca 2000, Bilir et al. 2005, Dall \& Schmidtobreick 2005).
Unfortunately, V568 Peg is too faint for radial velocity
measurements based on spectral observations by the 2-m telescope
at NAO Rozhen.

After reaching the best light curve solution we adjusted the
stellar temperatures $T_{1}$ and $T_{2}$ around the value $T_m$ by
the formulae (Kjurkchieva $\&$ Vasileva 2015)
\begin{equation}
T_{1}^{f}=T_{\rm {m}} + \frac{c \Delta T}{c+1}; \quad
T_{2}^{f}=T_{1}^{f}-\Delta T
\end{equation}
where the quantities $c=l_2/l_1$ (the ratio of the relative
luminosities of the stellar components) and $\Delta T=T_{m}-T_{2}$
are determined from the \textsc{PHOEBE} solution. In fact,
formulae (1) are consequence of the $T_m$ definition given
earlier.

Last fitting procedure was carried out for fixed $T_{1}^{f}$ and
$T_{2}^{f}$ to obtain the final and self-consistent solution.

\begin{figure}[!htb]
  \begin{center}
    \centering{\epsfig{file=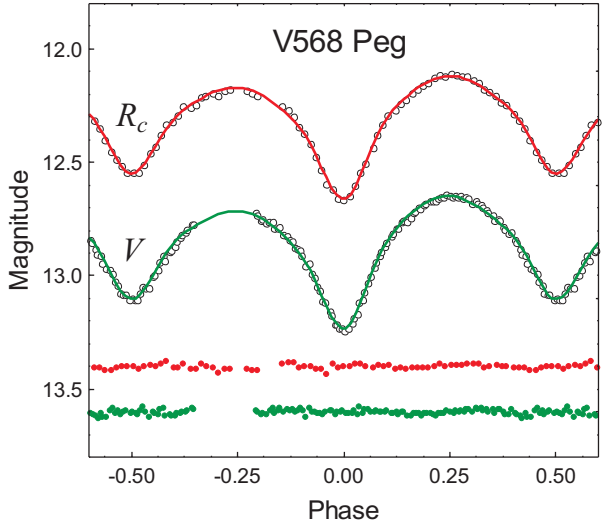, width=0.6\textwidth}}
    \caption[]{Top: the folded light curves of V568 Peg and
their fits; Bottom: the corresponding residuals (shifted
vertically to save space)}
    \label{fig1}
  \end{center}
\end{figure}

\begin{figure}[!htb]
  \begin{center}
    \centering{\epsfig{file=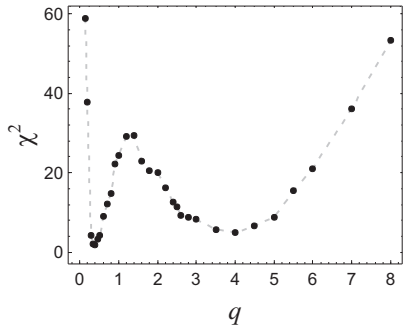, width=0.4\textwidth}}
    \caption[]{$q$-search curve}
    \label{fig2}
  \end{center}
\end{figure}

\textsc{PHOEBE} gives as output parameters the relative (volume)
radius $r_i=R_i/a$ of each component ($R_i$ is linear radius and
\emph{a} is orbital separation). One can determine the luminosity
ratio $c=L_2/L_1=l_2/l_1$ from the \textsc{PHOEBE} output
parameter $M_{bol}^2$ -- $M_{bol}^1$. The output potentials
$\Omega(L_1)$ and $\Omega(L_2)$ allowed to calculate the target
fill-out factor $f = [\Omega - \Omega(L_1)]/[\Omega(L_2) -
\Omega(L_1)]$.

We estimated the precision of the fitted parameters by the
procedure described in Dimitrov et al. (2017).

Table~2 contains the final values of the fitted stellar and spot
parameters and their uncertainties. Table 3 exhibits the
calculated parameters: $r_{1, 2}$, $f$ and $l_2/l_1$. Their errors
are determined from the uncertainties of fitted parameters. The
synthetic curves corresponding to the parameters of our light
curve solution are shown in Fig. 1 as continuous lines while Fig.
3 exhibits the 3D configuration of V568 Peg.

\begin{table}[htb]
  \begin{center}
  \caption{Fitted parameters of the best light curve solution}
\begin{scriptsize}
\begin{tabular}{cllccclcrccc}
  \hline\hline
$T_0$ - 2450000 & $P$         &  $i$         &  $q$     & $T_{1}^{f}$ & $T_{2}^{f}$  & $\Omega$ & $\beta$      & $\lambda$    & $\alpha$     & $\kappa$ &  source  \\
                & [d]         & [$^{\circ}$] &          & [K]         &   [K]        &          & [$^{\circ}$] & [$^{\circ}$] & [$^{\circ}$] &          &            \\
 \hline
6925.380188(24) & 0.247095(1) & 76.3(1)      & 0.494(1) & 5734(48)    & 5409(17)     & 2.786(2) &   70(1)      &   75(1)     &   22.0(2)     & 0.85(1)  & Paper I\\
6925.533810(14) & 0.247095(2) & 75.6(1)      & 0.403(3) & 4963(30)    & 4713(23)     & 2.6343(1)&   70(1)      &   75(1)     &   20.5(2)     & 0.88(1)  & this paper\\
    \hline
 \end{tabular}
\end{scriptsize}
  \label{table2}
  \end{center}
\end{table}

\begin{table}[htb]
  \begin{center}
  \caption{Calculated parameters}
\begin{scriptsize}
\begin{tabular}{ccccc}
  \hline \hline
$r_{1}$  & $r_{2}$  & $f$   & $l_{2}/l_{1}$ & source\\
 \hline
0.462(6) & 0.340(4) & 0.265 &   0.35        & Paper I\\
0.475(6) & 0.317(9) & 0.204 &   0.36        & this paper \\
 \hline
\end{tabular}
\end{scriptsize}
  \label{table3}
  \end{center}
\end{table}

\begin{figure}[!htb]
  \begin{center}
    \centering{\epsfig{file=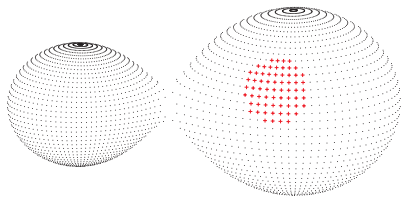, width=0.4\textwidth}}
    \caption[]{3D configuration of V568 Peg made using Binary
Maker 3 by Bradstreet and Steelman (2002).}
    \label{fig3}
  \end{center}
\end{figure}

\begin{figure}[!htb]
  \begin{center}
    \centering{\epsfig{file=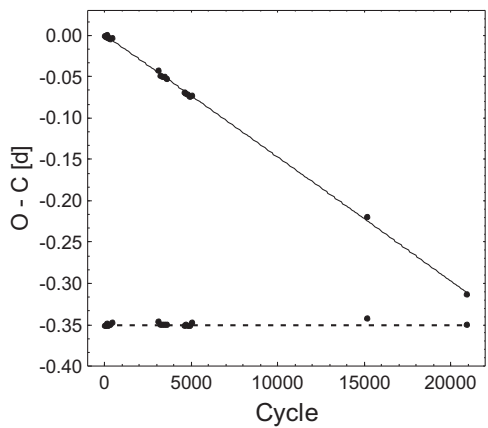, width=0.5\textwidth}}
    \caption[]{O-C diagram of V568 Peg: the three clusters of points correspond to WASP data while the last two points to our two solutions}
    \label{fig4}
  \end{center}
\end{figure}

\section*{4. Global parameters of V568 Peg}

The \emph{GAIA} distance of V568 Peg is 253 pc (Bailer-Jones et
al. 2018). It allowed us to calculate the target global parameters
by the following procedure.

(a) We obtained the target absolute magnitude $M_V$ = 5.884 mag by
the formula of distance modulus using its visual magnitude $V$ =
12.81 mag at quadrature (the extinction $A_V$=0.134 was estimated
according to Arenou et al. (1992)).

(b) The bolometric magnitude $M_b$ = 5.794 mag was calculated from
$M_V$ and the bolometric correction BC = -0.36 mag (Masana et al.
2006) corresponding to temperature 4900 K.

(c) The total luminosity $L$ = 0.537 L$_{\odot}$ was obtained from
$M_b$.

(d) The individual luminosities $L_1$ = 0.394 L$_{\odot}$ and
$L_2$ = 0.143 L$_{\odot}$ were calculated by $L$ and ratio
$c=L_2/L_1$ = 0.36 from the \textsc{PHOEBE} solution.

(e) The component radii $R_1$ = 0.851 R$_{\odot}$ and $R_2$ =
0.569 R$_{\odot}$ were determined from the individual luminosities
$L_i$ and temperatures $T_i$ (Table 2).

(f) The orbital axis $a$ = 1.796 R$_{\odot}$ was calculated from
the absolute radii $R_i$ and relative stellar radii $r_i$ (Table
3).

(g) The total mass $M$ = 1.272 M$_{\odot}$ was determined by the
third Kepler law based on the orbital axis $a$ and target period
$P$.

(k) The individual mases $M_1$ = 0.907 M$_{\odot}$ and $M_2$ =
0.365 M$_{\odot}$ were obtained from $M$ and the mass ratio $q$
(Table 2).

\section*{5. Analysis of the results}

The comparison of the parameter values of the new solution (Tables
2--3) with the previous one (Paper I) led to the following
results.

(1) The difference in inclination is negligible.

(2) There is small difference in relative component radii (1.5
$\%$ for the primary radius and 7 $\%$ for the secondary radius).

(3) The temperature differences of the components $\Delta T =
T_1^f-T_2^f$ are close while the component temperatures themselves
$T_1^f, T_2^f$ differ by around 700 K (Table 2). This is an
illustration of the known fact that the light curve solution is
strongly sensitive to $\Delta T$ but not to the individual
temperatures. Thus, the two solutions could be considered as
similar in temperature parameters. This is supported by almost the
same luminosity ratios of the two solutions (Table 3). We assume
that the new solution is more confident due to the determination
of its $T_m$ value by several dereddened color indices.

(4) The values of the mass ratio differ by 18 $\%$. We attributed
the difference to using of third light (of around 0.065) in the
previous solution (Paper I) because the new $V, R_c$ data can be
well-reproduced by the parameters of old solution including the
the same third light value. But we assume that the new solution is
more reasonable because it does not require an art third light
(invisible around the target). Moreover, the new $q$ value is more
confident because it is obtained by detailed $q$-search analysis.

(5) The biggest difference of the two solutions is the value of
the initial epoch $T_0$ (Table 2). It implies shorter period than
the known value. To check this supposition we determined times of
light minima of SWASP data (Butters et al. 2010). From the O--C
diagram (Fig. 4) we derived the period value of
0.2470800$\pm$0.0000003 d.


(6) The closeness of the spot positions and spot parameters of the
two solutions (Table 2) means existing of stable large cool spot
on the primary component during the last 4 years, i.e. during
around 5740 cycles.

\section*{6. Conclusion}

Our $V, R_c$ observations of V568 Peg revealed that the target is
an overcontact binary of A subtype with moderate fill-out factor.
Its components are K stars which undergo partial eclipses. The new
data allowed us to improve the target period while the \emph{GAIA}
distance provided a possibility to calculate the masses, radii and
luminosities of its components.

\section*{Acknowledgments}

The research was supported partly by projects DN08/20 and DM08/02
of Scientific Foundation of the Bulgarian Ministry of Education
and Science, project D01-157/28.08.2018 of the Bulgarian Ministry
of Education and science, as well as by projects RD-08-142 and
RD-08-112/2018 of Shumen University.

The authors are very grateful to the anonymous Referee for the
valuable notes and recommendations.

This work has made use of data from the European Space Agency
(ESA) mission Gaia (https://www.cosmos.esa.int/gaia), processed by
the Gaia Data Processing and Analysis Consortium (DPAC). Funding
for the DPAC has been provided by national institutions, in
particular the institutions participating in the Gaia Multilateral
Agreement.

\end{document}